# High-Yield Production and Transfer of Graphene Flakes Obtained by Anodic Bonding

Thomas Moldt[1], Axel Eckmann[2], Philipp Klar[1], Sergey V. Morozov[3], Alexander A. Zhukov[4], Kostya S. Novoselov[5], Cinzia Casiraghi*[1,2]

1 Physics Department, Free University Berlin, Germany; 2 School of Chemistry and Photon Science Institute, University of Manchester, UK; 3 Institute for Microelectronics Technology, Chernogolovka, Russia; 4 Manchester Centre for Mesoscience and Nanotechnology, University of Manchester, UK; and 5 School of Physics and Astronomy, University of Manchester, UK

email: cinzia.casiraghi@manchester.ac.uk

Abstract

We report large-yield production of graphene flakes on glass by anodic bonding. Under optimum conditions, we counted several tens of flakes with lateral size around 20-30 μm and few tens of flakes with larger size. 60-70% of the flakes have negligible D peak. We show that it is possible to easily transfer the flakes by wedging technique. The transfer on silicon does not damage graphene and lowers the doping. The charge mobility of the transferred flakes on silicon is of the order of 6000 cm$^2$/V s (at carrier concentration of $10^{12}$ cm$^{-2}$), which is typical for devices prepared on this substrate with exfoliated graphene.

KEYWORDS: graphene, transfer, anodic bonding, Raman Spectroscopy

Graphene is a two-dimensional hexagonal lattice of carbon atoms. Several graphene sheets stacked give ordinary three-dimensional graphite crystals. Graphene attracts enormous interest because of its unique properties.[1–7] Near-ballistic transport at room temperature and high mobility[6–11] make it a potential material for nanoelectronics,[12–16] especially for high frequency applications. Furthermore, its optical and mechanical properties are ideal for micro and nanomechanical systems, thin-film transistors, transparent and conductive composites and electrodes, and photonics.[17–22]

The most used technique to produce graphene flakes is based on the Micro-Mechanical Exfoliation (MME) of graphite. [2,23] This is a very simple and cheap method that requires some graphite flakes and scotch tape only. However, the graphene yield is very low: graphene flakes are rare, while few graphene layers and thick piece of graphite mostly cover the whole substrate. Thus, identification of graphene is time-consuming and relatively difficult, in particular when graphene is deposited on transparent substrates. Furthermore, the graphene flakes produced by micro-mechanical exfoliation of graphite have relatively small size, typically with lateral size of 10-20 μm. It is possible to strongly increase the graphene size by using some special cleaning treatment of the substrate: flakes up to 1 mm lateral size have been produced. However, the yield still remains very low: less than 3-4 large flakes per substrate are typically produced. Furthermore, often the large flakes are covered by bubbles[24] whose origin is still unknown.

Alternative techniques have been developed in order to produce graphene's wafers such as epitaxial growth on SiC[25] and chemical vapor deposition on metals.[26,27] They all require a complex, relatively expensive setup, and careful control of the deposition parameters in order to grow graphene of high quality.

Finally, a different approach is based on anodic bonding technique, typically used to bond borosilicate glass and silicon wafers.[28,29] Anodic bonding is achieved by pressing borosilicate glass on a silicon wafer at high temperatures (above 200ºC), while an high electrostatic field is applied perpendicular to the layers. Due to heating the $Na_2O$ impurities in the glass decompose in $Na^+$ and $O^{2-}$ ions. The $Na^+$ ions are lighter and have a higher mobility compared to the $O^{2-}$ ions. The polarity of the voltage is chosen so that the $Na^+$ ions move away from the silicon-glass interface to the back contact[30]. The $O^{2-}$ ions remain at the interface causing a strong electric field there, which allows bonding between silicon and glass. Covalent Si-O-Si bonds are formed at the interface. This method can be used also to deposit graphene, where graphene replaces the silicon in the original technique.[31,32] This method allows quick and cheap production of graphene layers at a high yield: under optimum parameters an area comparable with the size of the graphitic flake is all covered by graphene and few graphene layers, with typical size well above 10 μm, some up to 1 mm.[31] The flakes have been studied by Raman Spectroscopy, Atomic Force Microscopy, Scanning Tunneling Microscopy and transport.[31,32] Despite the simplicity of the setup, this method has been rarely adopted[31,32] mainly because the anodic bonding technique allows depositing graphene on substrates with relatively mobile ions.[32] Thus, the most used substrate has been borosilicate glass.[32] This substrate is useful to study the optical properties of graphene[33] but it is not suitable for transport measurements because of the higher complexity in the lithography process and lack of back-gate. Furthermore, the optical visibility of graphene on glass is extremely low,[34] so the flake is very difficult to spot on the substrate. However, by using the anodic bonding method the location of the flakes on glass is straightforward: after anodic bonding, the glass surface is no longer smooth, *i.e.* the area of the coverslip which was treated by high voltage and temperature becomes opaque, so it is well visible by eyes (Figure 1(a) in Supporting Information). Furthermore, this area is mostly covered by single-layers flakes among bilayers and very few thick layers. This makes graphene flakes produced by anodic bonding the perfect samples for optical spectroscopy and near-field measurements.

Previous works also report electrostatic deposition of graphene on oxidized silicon by applying a voltage well above 3 kV.[35-36] However, under these conditions, the control of the thickness is very difficult: few layers are generally deposited, and the quality of the flakes is very low, since the Raman spectrum shows a very intense D peak.[35] No information on the yield of single layers is reported. This method has been used for electrostatic printing of few graphene layers arrays and nanoribbons.[37-38]

In this work we show an extensive analysis of the properties of graphene produced by anodic bonding. Raman Spectroscopy has been used to identify single-layers graphene and to probe doping and disorder. The peaks have been fitted with a single Lorenzian lineshape and we analyzed the position (Pos), Full With at Half Maximum (FWHM) and intensity (I) of the G and 2D peaks (here intensity is the integrated area of the peak). We show that Anodic Bonding is a valid alternative to MME, since it allows producing high yield and defects-free graphene flakes with a very simple setup. We show that the flakes can be easily transferred with no damage to other substrates, such as silicon (Si/SiO$_x$), by the wedging technique.

**RESULTS AND DISCUSSION**

In the anodic bonding a single crystal flake of graphite is pressed on glass, a high voltage of 0.5 - 2 kV is applied between the graphite and a metal back contact, while heating the glass at about 200ºC for 10-20 minutes. In case of the positive electrode applied to the top contact, a negative charge concentration occurs in the glass at the side facing the positive electrode. Few-layers of graphite, including single layers, stick on the glass by electrostatic interaction. The anodic bonding is a simple technique because there are two deposition parameters only: temperature and voltage. Thus, in order to determine the optimum conditions to have high yield and high quality single layers graphene, we made several samples at different temperature (between 160 and 260ºC) and voltage (between 0.4 and 3 kV).

First, we investigated the samples by optical microscope. Figure 1(a) shows a sample obtained at 220ºC and 0.4kV: the substrate is mostly empty, the bonding efficiency is very low. Figure 1(b) shows a sample obtained at 220ºC and 0.9kV: the substrate is well covered by graphene flakes, few layers graphene are visible too. Figure 1(c) shows a sample obtained at 220ºC and 1.5kV: the large single layer sheets are broken into small flakes, with no defined edges. At higher voltage disruptive discharges through the glass can be observed: 3kV is the upper limit for the applied voltage, under our experimental conditions. Thus, we found that the bonding efficiency of graphene on glass is maximum between 0.6 and 1.2kV: moving to higher voltage strongly damages the largest flakes. Figure 1(d) shows a sample obtained at 260ºC and 0.9 kV: only few graphitic particles and thick flakes are visible on the glass surface, which is no longer smooth, but it shows circular spots and lines. Our results show that the bonding is starting to be efficient above 180ºC, but above 260ºC lot of thicker flakes and particles are visible. Thus, we can conclude that at low temperature, the mobility of ions in the glass is not high enough for achieving a strong electrostatic interaction between graphene and glass. In contrast, if the temperature is too high, then the efficiency of the bonding is high, allowing to bond even thick graphite to the glass and thinner flakes are damaged.

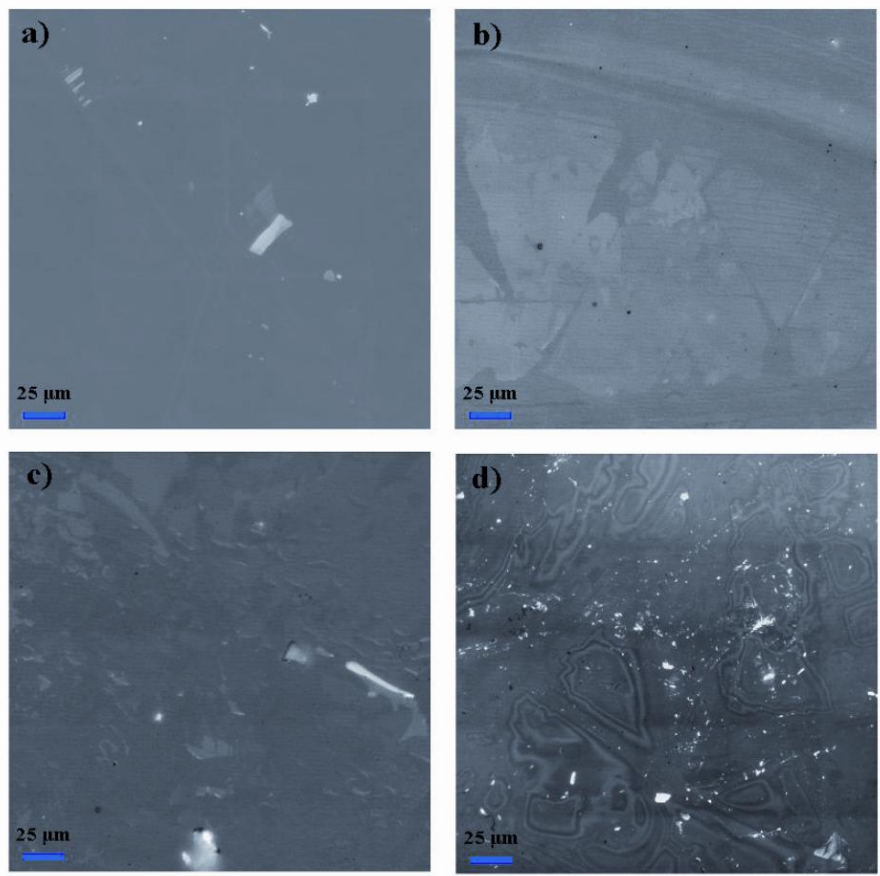

**Figure 1** Optical microscope pictures of the flakes deposited by anodic bonding under different conditions: (a) at 220ºC and 0.4 kV very few and small graphene flakes are visible, *i.e.* the bonding efficiency is very low; (b) at 220ºC and 0.9 kV several large graphene and bilayers flakes are deposited and good area coverage is achieved; (c) at 220ºC and 1.5 kV the bonding efficiency is high, but the sheets are broken into small flakes; (d) at 260ºC and 0.9 kV there are only thick flakes and particles.

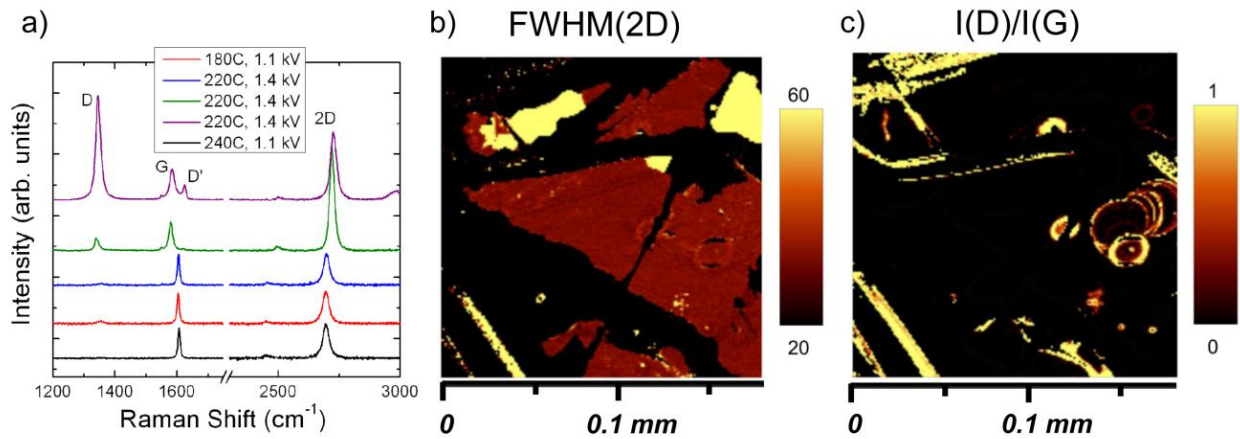

**Figure 2** (a) Raman spectra of the flakes obtained under different conditions; (b,c) Raman map of the FWHM of the 2D peak (the scal bar is in cm$^{-1}$) and intensity ratio between the D and G peak (scal bar is in arb. units) of a sample deposited at 0.9 kV and 220ºC. The flake in the center of Fig.2 (b), with lateral size of about 0.1 mm, is a graphene as indicated by the FWHM(2D) of about 30 cm$^{-1}$. The D peak is localized only at the edges of this flake. Other single layers flakes are visible.

Under our experimental conditions, we found that the highest yield of graphene is obtained in the range 180-240ºC and 0.6-1.2 kV. Under these conditions, we counted several tens of flakes with lateral size around 20-30 μm and few tens of flakes with larger size. Few flakes with lateral size of about 100 μm have been observed too. Note that the optimum voltage and temperature strongly depends on the type of substrate.[32]

Since the process involves high temperature and voltage, it is fundamental now to investigate the quality of the flakes. Figure 2 (a) shows the typical Raman spectra measured on flakes obtained under the range 180-240ºC and 1.1-1.4kV. First, we can note that the 2D peak is a single and sharp peak, which confirms that the flakes are single-layers.[39] Second, we can see that some of the Raman spectra show defect-activated peaks, D and D' peaks.[39] We found that the D peak intensity strongly depends on temperature and voltage: the higher these parameters, the higher the probability that the flake will have a strong D peak. Most of the single layers deposited at 0.6 kV and 220ºC do not show any D peak. For increasing voltage, the D peak starts to appear in some of the flakes: at 1.1 kV most of the flakes have D peak, although its intensity is usually up to 10-20% the G peak intensity. At higher voltage all the flakes have large D peak, sometimes disordered carbon is also observed.

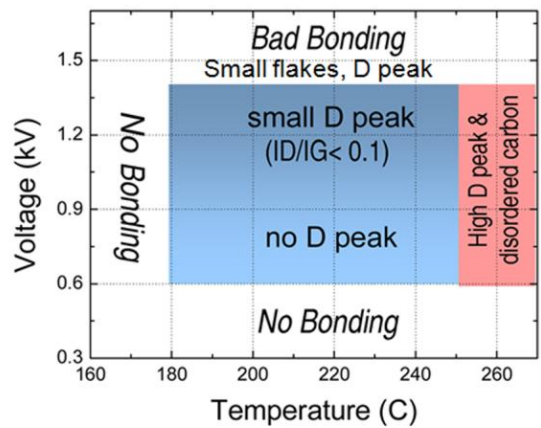

**Figure 3** Schematic of the properties of most of the flakes obtained with different deposition parameters with our anodic bonding setup.

Figure 2(b,c) shows a Raman map of the FWHM(2D) and intensity ratio between D and G peak, I(D)/I(G), of some flakes deposited under optimum conditions. A flake with lateral size of about 100 μm is visible. This is a single layer graphene, as indicated by its FWHM(2D) of about 30 cm$^{-1}$.[39] The D peak is visible only at the edges. Smaller single layers and bilayers are visible too.

Figure 3 gives a schematic overview of the quality of the graphene flakes obtained under different deposition parameters. Finally, we can observe that even in absence of a D peak, the Raman spectra show variations in the peaks positions and FWHM. This can be well attributed to doping. Doping is expected in those samples because there are charges involved in the Anodic Bonding method.[32] In order to confirm that the samples can be doped, we compared the Raman fit parameters of G and 2D peaks with the ones measured in pristine graphene on Si/SiO$_x$, and gated graphene,[40-42] Figure 4: a very good agreement in the variation of the G and 2D peak shape is observed. A high Pos(G) and large FWHM(G) corresponds to low doping, while the Pos(2D) can be used to distinguish between n- and p-doping.[40-42] Figure 4 shows that Anodic Bonding graphene can be doped and that the doping is p-type, as observed from Pos(2D) measured on the samples with high doping.

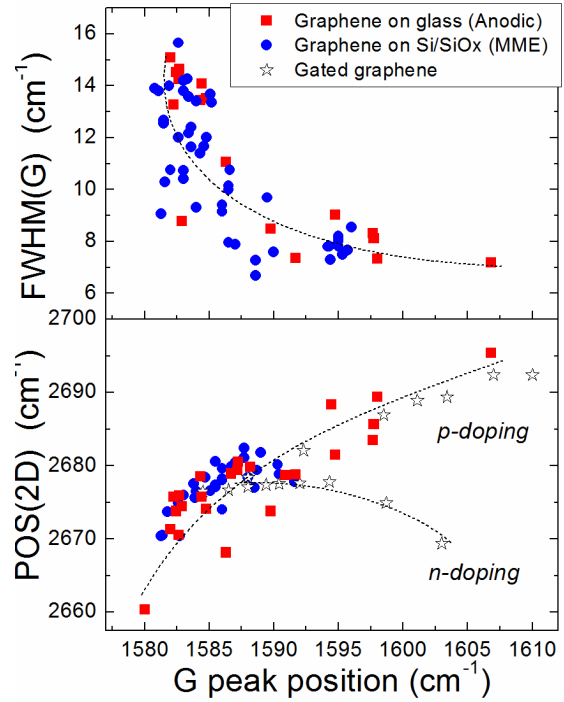

**Figure 4** Raman fit parameters for the G peak and 2D peak of graphene deposited on glass by anodic bonding, compared with the Raman fit parameters of exfoliated graphene deposited on Si/SiOx. The data of gated graphene are taken from Ref. 41. The dotted lines are only a guide for the eyes.

We transferred two graphene flakes deposited under the same conditions on two glass coverslip to: i) a new glass coverslip, in order to check if the doping is related to the glass substrate; ii) on Si/SiO$_x$, for transport measurement. Figure 5(a,b) shows a graphene sheet transferred from a coverslip to a new clean coverslip. Figure 5(c) shows the Raman spectrum before and after transfer. Note that there is no visible D peak after transfer. By fitting the Raman spectra, Fig. 5(c) we found that the Pos(G) decreased from 1607 to 1594 cm$^{-1}$, while FWHM(G) increased from 7 to 11 cm$^{-1}$. These variations show that after transfer the doping in graphene is strongly lowered. This further confirms that doping in anodic bonding graphene is mainly related to the charges used to deposit graphene on glass. However, the doping can not be completely removed by the transfer on a new substrate. Strain effect is also possible. Figure 5(d,e) shows a graphene sheet transferred from a coverslip to a clean Si/SiO$_x$ substrate. Figure 5(f) shows the Raman spectrum before and after transfer. Note that there is no visible D peak after transfer. By fitting the Raman spectra, we found that the Pos(G) decreased from 1591 to 1588 cm$^{-1}$, while FWHM(G) increased from 9 to 13 cm$^{-1}$. These variations show that after transfer the doping in graphene is strongly lowered, reaching the typical doping level observed for graphene deposited on Si/SiO$_x$ by MME.[40]

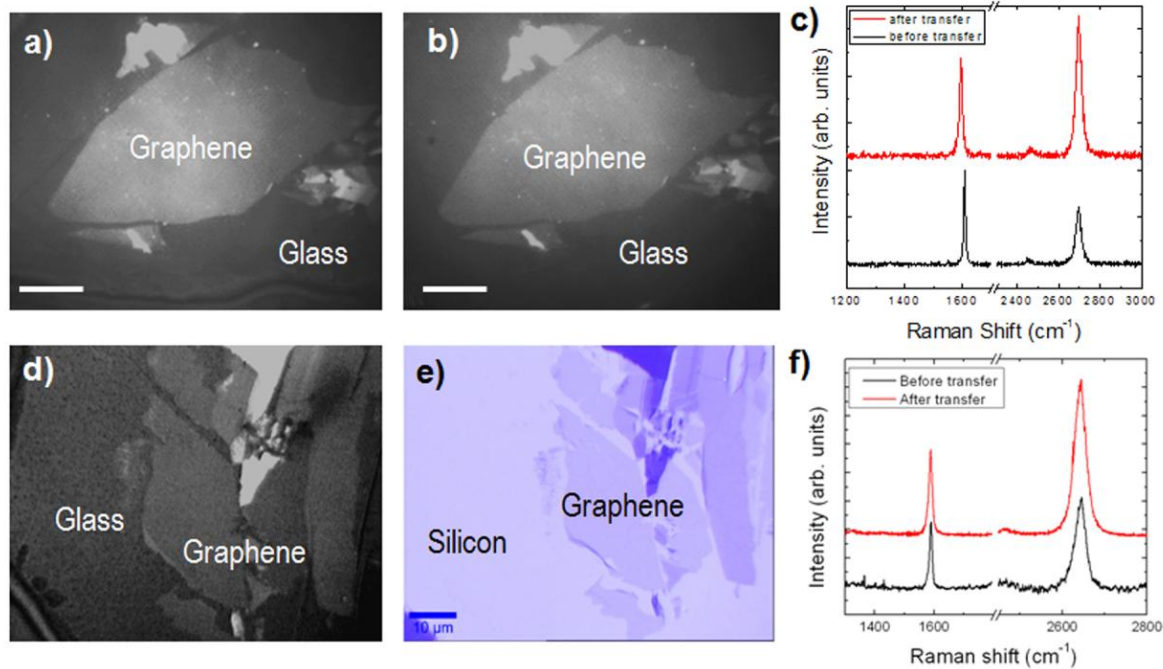

**Figure 5** Optical picture of graphene as deposited on glass (note the change in the topography of the coverslip, bottom of the Fig.); (b) Optical picture of the same sheet transferred to a new and clean coverslip; (c) Raman spectra of the flake before and after transfer. No D peak is visible after transfer. The scale bar in a) and b) is 10μm; (d) Optical picture of another as deposited graphene flake on glass; (e) Optical picture taken after transfering the flake on Si/SiO$_x$; (f) Raman spectra of the graphene sheet before and after transfer.

Our experiments show that the flakes produced by Anodic Bonding on glass can be transferred on other substrates without introducing defects. The transfer on a new substrate decreases the amount of doping in the graphene flake. This is independent on the type of substrate.

In order to finally check the quality of the flakes we prepared field effect transistor devices and measured their transport characteristics. The inset in Figure 6 shows one of our Hall bar mesa structure. After brief annealing at 250ºC in forming gas, the samples appear practically undoped (of the order of $10^{11} cm^{-2}$ *p*-doping). The field effect mobility extracted from the slope of the conductivity curve (Fig. 6, top panel) is of the order of 6000 $cm^2/Vs$ (at carrier concentration $10^{12} cm^{-2}$), which is typical for devices prepared on Si/SiOx.[6] Our measurements in magnetic field reveal the half-integer quantum Hall effect (Fig. 6, bottom panel), which is characteristic of exfoliated graphene devices.[6,7]

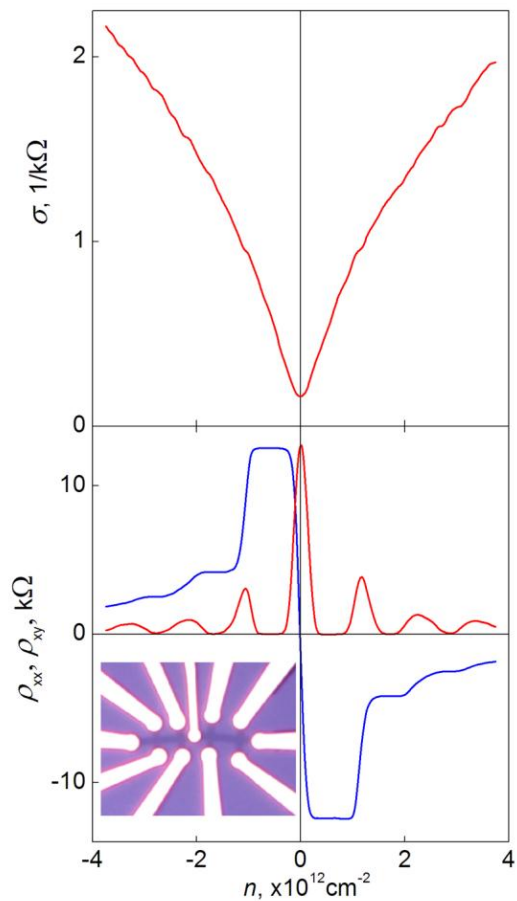

**Figure 6** Transport characteristics of a device prepared from a transferred graphene flake. Top panel: Conductivity as a function of carrier concentration. Bottom panel: Longitudinal (red) and transverse (Hall, blue) resistivity as a function of carrier concentration. Inset: a micrograph of our device. The scale is given by the width of the current lead (1 μm).

**CONCLUSIONS**

Micro-mechanical exfoliation of graphite is the most used method to produce graphene flakes on a substrate. Despite being simple and cheap, this technique can produce only few flakes. Furthermore, the identification of graphene can be very time-consuming when the single-layer is deposited on transparent substrates. Here, we show that it is possible to deposit large yield of graphene flakes on glass by anodic bonding techniques. Under optimum conditions, 60-70% of the flakes have negligible D peak. The flakes can be easily transferred onto other substrates, without damage, by wedging technique. The charge mobility measured after transfer on silicon is of the order of 6000 $cm^2/Vs$ (at carrier concentration $10^{12} cm^{-2}$), which is typical for devices prepared with exfoliated graphene on Si/SiO$_x$.

## METHODS

**Materials**: Single-crystal graphite flakes (National de Graphite) with size of 1.7 mm have been used to produce graphene. Few depositions have been performed with very large single-crystal graphite flakes, with size of 5 mm. The graphite flake is cleaved once using sticky tape in order to achieve a clean and fresh surface. The flake is then placed on a microscope coverslip, with thickness of 120 μm (Menzel-Gläser). The coverslip is cleaned before deposition by sonication in acetone and then isopropanol.

**Anodic bonding setup**: this is composed by a grounded metal block used as back electrode and can be heated up to 300°C using a temperature feedback controlled heating plate. The glass coverslip is placed on the grounded electrode. The top electrode, a cylindrical metal rod with a diameter of 2 mm, mounted vertically above the back gate, is pressed on the graphite flake, while applying a DC voltage for 20-30 minutes. The setup allows DC voltages of up to 10kV. After the deposition, thick graphite material is removed from the coverslip by using sticky tape.

**Monochromatic filter**: the contrast of a graphene sheet on glass illuminated in reflection mode is 7%. [43] However, the flakes were hardly visible under the microscope. We found that it is possible to strongly increase the contrast of the flake by converting the RGB image into a monochromatic image.

**Transfer and transport**: the graphene flakes produced by anodic bonding have been transferred to other substrates by using the wedging technique.[4] We transferred graphene flakes from the coverslip to a silicon substrate covered with 90 nm silicon oxide (IDB Technology) for transport measurements. Electron beam lithography and e-beam evaporation were used to prepare a set of contacts (5nm Ti/50nm Au). A Hall bar mesa structure has been prepared by reactive plasma etching.

**Raman Spectroscopy**: we used a WITEC alpha300 Raman Spectrometer, equipped with 488, 514 and 633 nm laser lines. The laser power was kept as low as 500 mW in order to avoid damage by laser heating. The spectral resolution is 2-3 cm$^{-1}$. The instrument is equipped with a piezostage, which allows doing Raman mapping with spatial resolution down to 10 nm. Further measurements have been taken with a HORIBA XploRA Confocal Raman Spectrometer, equipped with 532 nm laser wavelength. The theory of the Raman spectrum of graphene is described in the Supporting information.

**Acknowledgments**: the authors thank S. Reich for the use of the XploRA Raman spectrometer and F. Mauri for useful discussions. This work is funded by the Alexander von Humboldt Foundation in the framework of the Sofja Kovalevskaja Award, endowed by the Federal Ministry for Education and Research of Germany.

**Supporting Information Available:** some pictures of the flakes, details on the monochromatic filter, transfer technique and Raman spectroscopy background are available free of charge *via* the Internet at http://pubs.acs.org.